\documentclass[aps,prb,reprint,groupedaddress]{revtex4-2}
\usepackage{graphicx}
\usepackage{dcolumn}
\usepackage{bm}

\usepackage[utf8]{inputenc}
\usepackage[T1]{fontenc}
\usepackage{booktabs, array, mathptmx, float, tabularx, booktabs, lipsum, amsmath,multirow}
\usepackage{siunitx, xcolor}
\usepackage[version=4]{mhchem}
\graphicspath{{figs/}{figsgaoerb/}} 
\usepackage[colorlinks,linkcolor=blue,anchorcolor=blue,citecolor=blue]{hyperref}

\begin{document}


\title{Doping evolution of the normal state magnetic excitations in pressurized La$_3$Ni$_2$O$_7$}


\author{Hai-Yang Zhang}
\author{Yu-Jie Bai, Fan-Jie Kong, Xiu-Qiang Wu, Yu-Heng Xing}
\author{Ning Xu}
\affiliation{Department of Physics, Yancheng Institute of Technology, Yancheng 224051, China}


\date{\today}

\begin{abstract}
The doping evolution behaviors of the normal state magnetic excitations (MEs) of the nickelate La$_3$Ni$_2$O$_7$ are theoretically studied in this paper.  For a filling of $n=3.0$ which corresponds roughly to the material which realizes the superconductivity of about $80$ K under moderately high pressures above $14$ GPa, the MEs exhibit a square-like pattern centered at $(0,0)$ which originates from the intrapocket particle-hole scatterings. Furthermore, it was found that the MEs show very strong modulations on the interlayer momentum $q_{z}$. With the increasing of $q_{z}$, the patterns of the MEs change dramatically. In the large $q_{z}$ regime, they turn to be ruled by the interpocket excitation modes with significantly larger intensity which may play a vital role in the superconducting pairing. This hidden feature of the MEs was not revealed by previous studies. The established features of the MEs are found to be very robust against doping. They persist in the wide hole- or electron-doping regime around $n=3.0$. However, in the heavily electron-doped regime, a Lifshitz transition will occur. Consequently, the behaviors of the MEs change qualitatively across this transition. With the absence of the hole pocket, we predict that a strong tendency toward the spin-density-wave (SDW) order will develop due to the perfect nesting between the electron and the hole pockets at $n=4.0$. Thus, we have investigated the normal state behaviors of the MEs and their doping evolutions which may shed light on the mechanism of superconductivity in pressurized La$_3$Ni$_2$O$_7$. 
\end{abstract}


\maketitle

\section{The introduction}
The newly discovered superconductivity with a transition temperature above $80$ K in the moderately pressurized La$_3$Ni$_2$O$_7$ has attracted much attention in condensed matter community~\cite{SunNat}. Zero resistance was  achieved in single~\cite{zeroR1,zeroR2,zeroR3,zeroR4} and polycrystalline~\cite{zeroRpolycrystal} samples by using the liquid pressure medium. It was found that, generally, the emergence of superconductivity is accompanied by the lattice structure transition driven by the applied pressure~\cite{zeroR1,zeroR4}. According to a recent report, the superconductivity persists up to $90$ GPa with a right-triangle shaped phase region~\cite{zeroR4}. Although intensive studies have been carried out, the mechanism of superconductivity is still a hotly debated issue~\cite{Yaotbmodel,BilayerSu,dandsFLEX,LuzhongyiSpm,elementSubWu,DMRGzhanggm,WenOptical,EreminRPAandpairing,Hujiangping0.5Pi,DagottoPairing,Feshreso,Yangyifeng,YaodaoXtJ,KunJiangB1gpairing,zhoutaoPRB,YYFZGMZhangfuchun,LuChenWuCprl,WuCongEgorb,WangFaCPLpairing,WangqianghuaPrb,WengZhengYuGauge,zhangyangNC,XiangTaoDwave,ZhangyangDagottoSwave,ZhangYHtypeII,YangfanPRLOdefi}. The density functional theory (DFT) calculations reveal that the low energy electronic bands of La$_3$Ni$_2$O$_7$ are ruled by the Ni-$3d_{z^{2}}$ and Ni-$3d_{x^2-y^2}$ orbitals and the superconductivity shows up with the presence of a hole pocket mainly of the Ni-$3d_{z^{2}}$ orbital character around the Fermi level~\cite{zeroR1}. Both the band structure and the topology of the Fermi surfaces obtained by DFT are recently confirmed by the ARPES measurements at ambient pressure~\cite{ZXJNCarpes,ARPEScpl}. Due to the strong hybridization between the Ni-$3d_{z^{2}}$ orbital and the apical O-$p_{z}$ orbital, the interlayer antiferromagnetic coupling of the $3d_{z^{2}}$ orbitals was proposed to be important~\cite{ZhangYHtypeII,YYFZGMZhangfuchun,Yangyifeng}. So far, there have been accumulated evidences that a SDW order may develop around $150$ K in the parental compounds~\cite{FengRIXS,uSR,uSRsplitDW,ChenxianhuiNMR}. In this way, the external pressure seems to play the similar role of charge carrier doping in cuprates to suppress the magnetic order and give birth to superconductivity. This proximity of the magnetic phase to superconductivity resembles that of many other unconventional superconductors, such as copper oxides~\cite{CupratesShenZhiXunRMP,CupratesZongshu} and iron-based pnictides~\cite{ScalapinoRMP}. As well known, it was believed that the magnetic fluctuations play a vital role in pairing of these materials~\cite{ScalapinoRMP}. To date, several candidates for the mechanism of superconductivity in pressurized La$_3$Ni$_2$O$_7$ had been theoretically proposed based on the magnetic-mediated interaction scenarios~\cite{dandsFLEX,EreminRPAandpairing,LuChenWuCprl,WangqianghuaPrb,YYFZGMZhangfuchun,zhangyangNC}. However, the origin of superconductivity is still unsettled. Thus, the investigation of the normal state MEs is highly desired as it may reveal the intimate relation between magnetic fluctuations and superconductivity in pressurized La$_3$Ni$_2$O$_7$.

In this paper, adopting the previously proposed bilayer-two-orbital model of Ref.~\cite{Yaotbmodel}, we theoretically studied the normal state behaviors of the MEs in the nickelate La$_3$Ni$_2$O$_7$. As indicated by the DFT calculations, the band structures of La$_3$Ni$_2$O$_7$ exhibit remarkable dependence on pressure. The hole pocket is slightly below the Fermi level at ambient pressure, however, it approaches and finally intersects with the Fermi level in the high pressure regime~\cite{SunNat}. That is, the Lifshitz transition will develop, driven by the applied pressure. Besides this, it is unclear how pressure affects the electronic energy bands in details. Here, the doping evolution behaviors of the MEs were studied for the pressurized La$_3$Ni$_2$O$_7$. We hope some of the results may be applied to this material.  Firstly, we focus on the case with a filling number of $n=3.0$ which corresponds roughly to the material which realizes the superconductivity of about $80$ K under a pressure of about $14$ GPa~\cite{SunNat}. It was found that the behaviors of the MEs have very strong dependence on the vertical momentum $q_{z}$ associated with the displacement between the upper and the lower nickel-oxygen layers. With $q_{z}$ near zero for which the interlayer spins align parallelly, the MEs exhibit a square-like pattern centered at $(0,0)$ which originates mainly from the intrapocket particle-hole scatterings of the $\beta$ band. With the gradually increasing of $q_{z}$, the MEs change dramatically. Specifically, for large $q_{z}$ near $\pi$, they turn to be dominated by two kinds of modes associated with the interpocket particle-hole scatterings. One of which originates from the particle-hole scatterings between the $\alpha$ and the $\beta$ pockets. While the other arises from the scatterings between the $\beta$ and the $\gamma$ pockets. These two modes have larger intensity compared to those with small $q_{z}$. Thus, they are expected to play more prominent roles if the pairing of electrons is mediated by magnetic fluctuations. Secondly, we find that these features of the MEs are very robust in the wide doping region around $n=3.0$. The patterns of the MEs and their energy and $q_{z}$ dependence keep qualitatively unchanged regardless of whether the system is hole- or electron-doped. Finally, for the heavily electron-doped system, a Lifshitz transition will occur. We find that the behaviors of the MEs will change qualitatively across this transition. When the hole $\gamma$ pocket is absent, a strong tendency toward the SDW order develops due to the well-nesting behavior between the $\alpha$ and the $\beta$ pockets. Thus, we have studied extensively the normal state behaviors of the MEs across a wide doping regime, which may shed light on the mechanism of superconductivity in pressurized La$_3$Ni$_2$O$_7$.

\section{The model and formulas}
We adopt the bilayer-two-orbital Hamiltonian $H=H_{0}+H_{I}$ to carry out the calculations, where the tight-binding part of the Hamiltonian reads $H_{0}=\sum_{kab\sigma}^{LJ}\epsilon_{ab}^{LJ}(k)C^{+}_{La\sigma}(k)C_{Jb\sigma}(k)$~\cite{Yaotbmodel} with $L$, $J$ the layer indices, while $a$, $b$ and $\sigma$ the orbital and spin indices, respectively. The kinetic energy terms are $\epsilon_{xx/zz}^{LL}(k)=-2t_{1}^{x/z}\gamma_{k}-4t_{2}^{x/z}\gamma_{k}^{'}+
\epsilon^{x/z}-\mu$, $\epsilon_{xz}^{LL}(k)=2t_{3}^{xz}\gamma_{k}^{''}$, $\epsilon_{xx/zz}^{L\bar{L}}(k)=2t_{\perp}^{x/z}$ , $\epsilon_{xz}^{L\bar{L}}(k)=2t_{4}^{xz}\gamma_{k}^{''}$, with $\gamma_{k}=$cos $k_x+$cos $k_y$, $\gamma_{k}^{'}=$cos $k_x$cos $k_y$ and $\gamma_{k}^{''}=$cos $k_x-$cos $k_y$, while $x$ and $z$ denote the Ni-$3d_{x^2-y^2}$ and Ni-$3d_{z^2}$ orbital, respectively. The hopping integrals are $t_{1}^{x}=-0.483$, $t_{1}^{z}=-0.110$, $t_{2}^{x}=0.069$, $t_{2}^{z}=-0.017$, $t_{3}^{xz}=0.239$, $t_{\perp}^{x}=0.005$, $t_{\perp}^{z}=-0.635$, $t_{4}^{xz}=-0.034$, $\epsilon^{x}=0.776$ and $\epsilon^{z}=0.409$ as listed in Ref~\cite{Yaotbmodel}.

The interacting part of the Hamiltonian $H_{I}$ is given by,
\begin{eqnarray}
	H_{I}&=&U\sum_{L,i,a} n_{Lia\uparrow}n_{Lia\downarrow}+U^{'}\sum_{L,i,a<b}n_{Lia}n_{Lib} \nonumber\\
	&+&J\sum_{L,i,a<b}
	C^{+}_{Lia\sigma}C^{+}_{Lib\sigma^{'}}C_{Lia\sigma^{'}}C_{Lib\sigma} \nonumber\\
	&+&J^{'}\sum_{L,i,a\neq b} C^{+}_{Lia\uparrow}C^{+}_{Lia\downarrow}
	C_{Lib\downarrow}C_{Lib\uparrow}.
	\label{eqn.1}
\end{eqnarray}
Where $n_{Lia}=n_{Lia\uparrow}+n_{Lia\downarrow}$ is the occupation number of the orbital $a$ electron at site $i$ of layer $L$. $U$, $U'$, $J$, $J'$ are the
coefficients of the intraorbital interaction, interorbital
interaction, Hund coupling and pair hopping terms, respectively.
The constrains $U=U^{'}+J+J^{'}$ and $J=J^{'}$ are used throughout the paper as required by the
spatial and spin rotational symmetry.

We study the normal state MEs of La$_3$Ni$_{2}$O$_{7}$ within the RPA approximation which had been used to study the MEs of various unconventional superconductors~\cite{RPA1,RPA2,RPA3,RPA4,RPA5,RPA6,RPA7,RPA8,RPA9,RPA10,RPA11,RPA12,RPA13,RPA14}.
It had been justified by these studies that the main features of the MEs can be qualitatively captured by this method. In the present model, the single particle Green's function can be defined as $\hat{G}_{ab\sigma}^{LJ}(k,\tau)=-\langle T[C_{ La\sigma}(k,\tau)C_{Jb\sigma}^{+}(k,0)]\rangle$. Due to the spin $SU(2)$ symmetry of the model, we restricted ourself to the transverse spin susceptibility which is defined as
$\hat{\mathcal{X}}_{ab,cd}^{LJ,KM}(q,\tau)=\langle T[S^{-}_{La,Jb}(q,\tau)S_{Kc,Md}^{+}(q,0)]\rangle$, where $S^{-}_{La,Jb}(q)=\frac{1}{\sqrt{N}}\sum_{k}C_{La\downarrow}^{+}(k)C_{Jb\uparrow}(k+q)$ and $S^{+}_{La,Jb}(q)=\frac{1}{\sqrt{N}}\sum_{k}C_{La\uparrow}^{+}(k+q)C_{Jb\downarrow}(k)$ are the spin lowering and rising operators, respectively, with $N$ the number of the lattice sites.
The bare spin susceptibility can be expressed as  $\hat{\chi}^{LJ,KM}_{ab,cd}(q,\tau)=-\frac{1}{N}\sum_{k}G_{da\downarrow}^{ML}(k,-\tau)G_{bc\uparrow}
^{JK}(k+q,\tau)$.
After the fourier transformation, we get $\hat{\chi}_{ab,cd}^{LJ,KM}(q,i\omega_{n})=-\frac{1}{N\beta}\sum_{k,i\omega_{m}}G_{da\downarrow}
^{ML}(k,i\omega_{m})G_{bc\uparrow}^{JK}(k+q,i\omega_{m}+i\omega_{n})$,
where $\beta=\frac{1}{T}$ is the inverse of the temperature $T$.

Decomposed into the spin channel, the nonzero elements of the interaction matrix read as $\hat{U}_{aa,aa}^{LL,LL}(q)=U$, $\hat{U}_{aa,bb(a\neq b)}^{LL,LL}(q)=
J$, $\hat{U}_{ab,ba(a\neq b)}^{LL,LL}(q)=U^{'}$ and $\hat{U}_{ab,ab(a\neq b)}^{LL,LL}(q)=J^{'}$. In this way, the RPA spin susceptibility can be written as $\hat{\mathcal{X}}(q,i\omega_{n})=\hat{\chi}(q,i\omega_{n})(\hat{I}-\hat{U}(q)\hat{\chi}
(q,i\omega_{n}))^{-1}$. After the analytic continuation, the retarded spin susceptibility $\hat{\mathcal{X}}^{R}(q,\omega)=\hat{\mathcal{X}}(q,\omega+i\Gamma)$ can be obtained, where $\Gamma$ is the broadening factor which is set to be $0.02$ in this paper. Actually,
for the double-layer system La$_3$Ni$_2$O$_7$, the imaginary part of the spin susceptibility can be decomposed into the intralayer and interlayer components as $\mathcal{X}^{''}_{S}(q,\omega)=\sum_{Lab}$Im$(\hat{\mathcal{X}}^{R})_{aa,bb}^{LL,LL}(q,\omega)$ and $\mathcal{X}^{''}_{D}(q,\omega)=\sum_{Lab}$Im$(\hat{\mathcal{X}}^{R})_{aa,bb}^{LL,\bar{L}\bar{L}}
(q,\omega)$, respectively. Considering the displacement between the upper and the lower layers, there should exist $q_{z}$ modulation for the MEs. Taking this into account, we can get the spectra of the MEs through $\mathcal{X}^{''}(q,q_{z},\omega)=\mathcal{X}^{''}_{S}(q,\omega)+
\mathcal{X}^{''}_{D}(q,\omega)
\cos q_{z}$, here in this paper we have set the displacement between the upper and the lower layers to be the unit length.

For simplicity, the rigid band approximation is used to study the doping evolution behaviors of the normal state MEs of the nickelate La$_{3}$Ni$_{2}$O$_{7}$. Throughout the paper, $T$ is set to be $0.001$ and $J=U/4$ is assumed. The energies are all in unit of eV. All the calculations are carried out for a $256\times256$ lattice.

\section{The numerical results}
\subsection{Magnetic excitations for the system with $n=3.0$}
\begin{figure}
	\centering\includegraphics[width=0.45\textwidth]{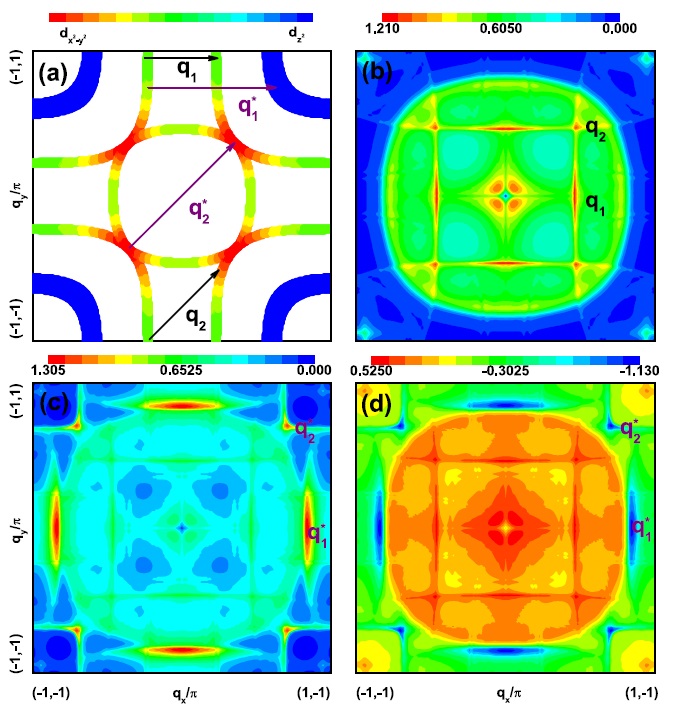}
	\caption{(Color online) Panel (a) is the Fermi surfaces and its orbital compositions for $n=3.0$. (b), (c) and (d) are the momentum distribution of the total, intralayer and interlayer magnetic spectra, respectively. The corresponding parameters are $U=1.1$, $\omega=0.02$, and $q_{z}=0$.}
	\label{U1.1OM0.02}
\end{figure}

The low energy MEs are tightly related to the topology of the Fermi surfaces. In panel (a) of Fig.~\ref{U1.1OM0.02}, we show the momentum distribution of the low energy electronic states within an energy window of $0.08$ eV around the Fermi energy and their dominated orbital components. The electron $\alpha$ and hole $\beta$ pockets are centered at $(0,0)$ and $(\pi,\pi)$ in the Brillouin zone, respectively. They are mainly composed of the $3d_{x^2-y^2}$ orbital, especially around the diagonal region of the Fermi surfaces. In contrast, the hole $\gamma$ pocket centered at $(\pi,\pi)$ are fully dominated by the $3d_{z^2}$ orbital. Meanwhile, it can be seen from panel (a) that the $\gamma$ pocket is much thicker than the $\alpha$ and $\beta$ ones, which indicates that it has larger density of states (DOS) around the Fermi level, partially due to its approach to the band edge.

In panel (b) of Fig.~\ref{U1.1OM0.02}, we show the momentum dependence of the MEs for $U=1.1$, $\omega=0.02$ and $q_{z}=0$. The MEs show ferromagnetic (FM) correlations as indicated by the four peaks around $(0,0)$. By decomposition of the spin susceptibility into the orbital channels, it can be confirmed that these FM MEs arise mainly from the $3d_{z^2}$ orbital. The contribution of the $3d_{x^2-y^2}$ orbital to these FM correlations are negligibly small. It was further found that these FM modes will be suppressed with the increasing of $U$. Away from $(0,0)$, there exist four peaks denoted by $q_{2}$ which reside along the diagonal direction and form the corners of a square. On the four edges of the square, other modes emerge with the characteristic momentum $q_1$, as shown in Fig.~\ref{U1.1OM0.02} (b). As indicated by the black arrows in Fig.~\ref{U1.1OM0.02} (a), both the $q_1$ and $q_2$ modes are attributed to the particle-hole scatterings of the $\beta$ band, due to the well-nested straight segments of the $\beta$ pocket around $(0,\pm\pi)$ and $(\pm\pi,0)$. This square-like pattern had been established by previous theoretical studies ~\cite{Yaotbmodel,EreminRPAandpairing}, although where only the real part of the zero energy spin susceptibility was concerned. To reveal the possible interlayer modulations of the MEs, we calculated the intralayer and the interlayer magnetic spectra of $\mathcal{X}^{''}_{S}$ and $\mathcal{X}^{''}_{D}$, respectively. The results are shown in panels (c) and (d) of Fig.~\ref{U1.1OM0.02}, correspondingly. Clearly, the square-like pattern remains for both the intralayer and the interlayer MEs. However, the most prominent features are four additional peaks around $(\pi,\pi)$ and four rod-like structure around $(\pm\pi,0)$ and $(0,\pm\pi)$, which are denoted as $q^{*}_{2}$ and $q^{*}_{1}$, respectively. As shown in Fig.~\ref{U1.1OM0.02} (c) and (d), comparing to that of the square-like pattern, these new MEs show significantly larger intensity. Most importantly, both the $q^{*}_{1}$ and $q^{*}_{2}$ modes show anti-phase relation between the intralayer and the interlayer spin susceptibility, which reflects the fact that there exists strong interlayer antiferromagnetic correlations. Thus, it can be expected that the MEs will exhibit strong $q_{z}$ dependence.

\begin{figure}
	\centering\includegraphics[width=0.45\textwidth]{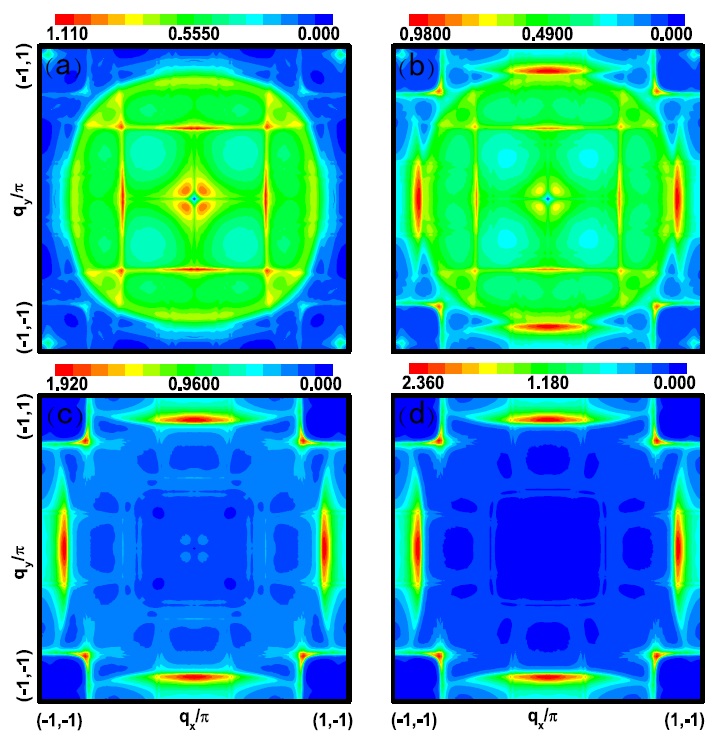}
	\caption{(Color online) $q_{z}$ evolution of the MEs for $U=1.1$, $\omega=0.02$. Panels (a), (b), (c) and (d) are the MEs for $q_{z}=0.2\pi$, $0.4\pi$, $0.7\pi$ and $\pi$, respectively.}
	\label{U1.1OM0.02qz}
\end{figure}

In Fig.~\ref{U1.1OM0.02qz}, for illustration, we show this $q_{z}$ evolution of the MEs for $U=1.1$ and $\omega=0.02$. Their patterns for $q_{z}=0.2\pi$, $0.4\pi$, $0.7\pi$ and $\pi$ are shown in panels (a), (b), (c) and (d), respectively. Clearly, with the increasing of $q_{z}$, the structure of the MEs change dramatically. For small $q_{z}$, the MEs are still dominated by the square-like pattern which arises from the intraband particle-hole excitations of the $\beta$ pocket. When $q_{z}$ increases to be around $0.4\pi$, two new $q^{*}_{1}$ and $q^{*}_{2}$ modes emerge, and they coexist with the original $q_1$ and $q_{2}$ modes. With further increasing of $q_{z}$, the $q_{1}$ and $q_{2}$ modes decrease in intensity gradually, and they disappear in the large $q_{z}$ regime. The resultant MEs are fully dominated by the $q^{*}_{1}$ and $q^{*}_{2}$ modes, as depicted in panels (c) and (d) of Fig.~\ref{U1.1OM0.02qz}. In contrast to the MEs with small $q_{z}$, where the dominate features are attributed to the intraband particle-hole excitations. For large $q_{z}$, they turn to be dominated by the interband particle-hole excitations. Specifically, the $q^{*}_{1}$ mode arises from the particle-hole excitations between the $\beta$ and the $\gamma$ pockets, while the $q^{*}_{2}$ mode mainly originates from the particle-hole scatterings between the $\beta$ and the $\alpha$ pockets. The typical scattering processes are shown in panel (a) of Fig.~\ref{U1.1OM0.02}. The above mentioned hidden $q_{z}$ dependent structures of the MEs had not been revealed by the previous studies~\cite{Yaotbmodel,EreminRPAandpairing}. It is interesting to notice that the MEs pattern for large $q_{z}$ near $\pi$ shares quite similar structure with that of the previously established SDW interaction $V_{SDW}(q)$ which was proposed to be responsible for the emergence of superconductivity~\cite{WangqianghuaPrb}. Here, we identify that the $q_{1}^{*}$ and $q_{2}^{*}$ modes are associated with the interlayer antiferromagnetic correlations. They shall play an important role if the magnetic fluctuation is regarded as the pairing glue.

\begin{figure}
	\centering\includegraphics[width=0.45\textwidth]{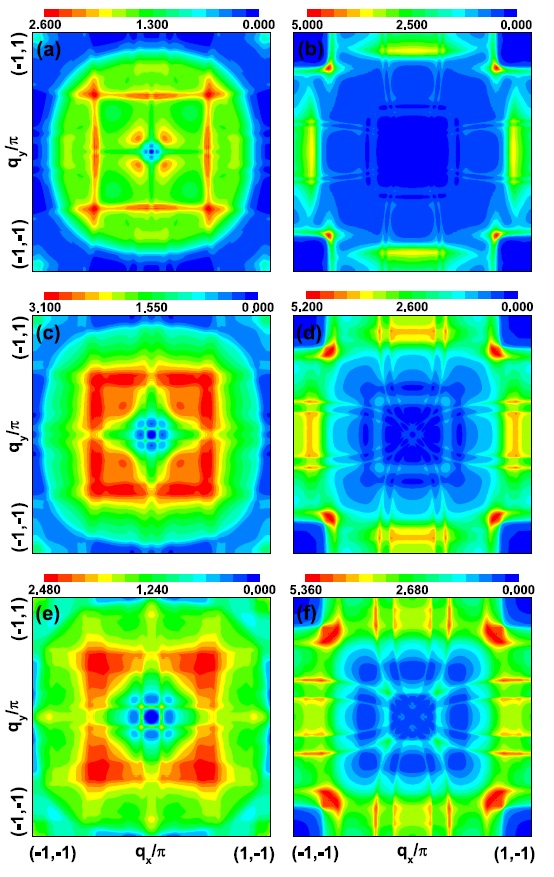}
	\caption{(Color online) Energy evolution of the MEs. Panels (a) and (b) are the MEs for $q_{z}=0$ and $q_{z}=\pi$, respectively, at a given energy of $\omega=0.06$. Panels (c), (d) and (e), (f) are the same with (a), (b) but for $\omega=0.12$ and $0.2$, respectively.}
	\label{MEomEVO}
\end{figure}

Now we turn to the energy evolution of the MEs. The typical patterns of the MEs are shown in Fig.~\ref{MEomEVO} for $\omega=0.06$, $0.12$ and $0.2$. For small $q_{z}$, with the increasing of energy, each of the $q_{1}$ modes splits into two parts which move gradually toward the location of the $q_2$ modes and finally merge with them in the high energy regime, giving rise to the significantly broadened peaks in momentum space. Generally, the broadening of the peaks of the MEs can be expected due to the much more available particle-hole pairs when the energy becomes large. For large $q_{z}$ near $\pi$, the $q_{2}^{*}$ modes persist up to the high energy regime. Also, their peaks broaden steadily with the increasing of energy. In contrast, as shown in panels (d) and (f) of Fig.~\ref{MEomEVO}, the $q_{1}^{*}$ modes split gradually into multiple peaks. This splitting of the $q_{1}$ and $q_{1}^{*}$ modes with energy can be understood by analyzing the nesting behaviors of the corresponding constant energy contours (CECs). But here we only focus on the main features of the MEs.   

\begin{figure}
	\centering\includegraphics[width=0.45\textwidth]{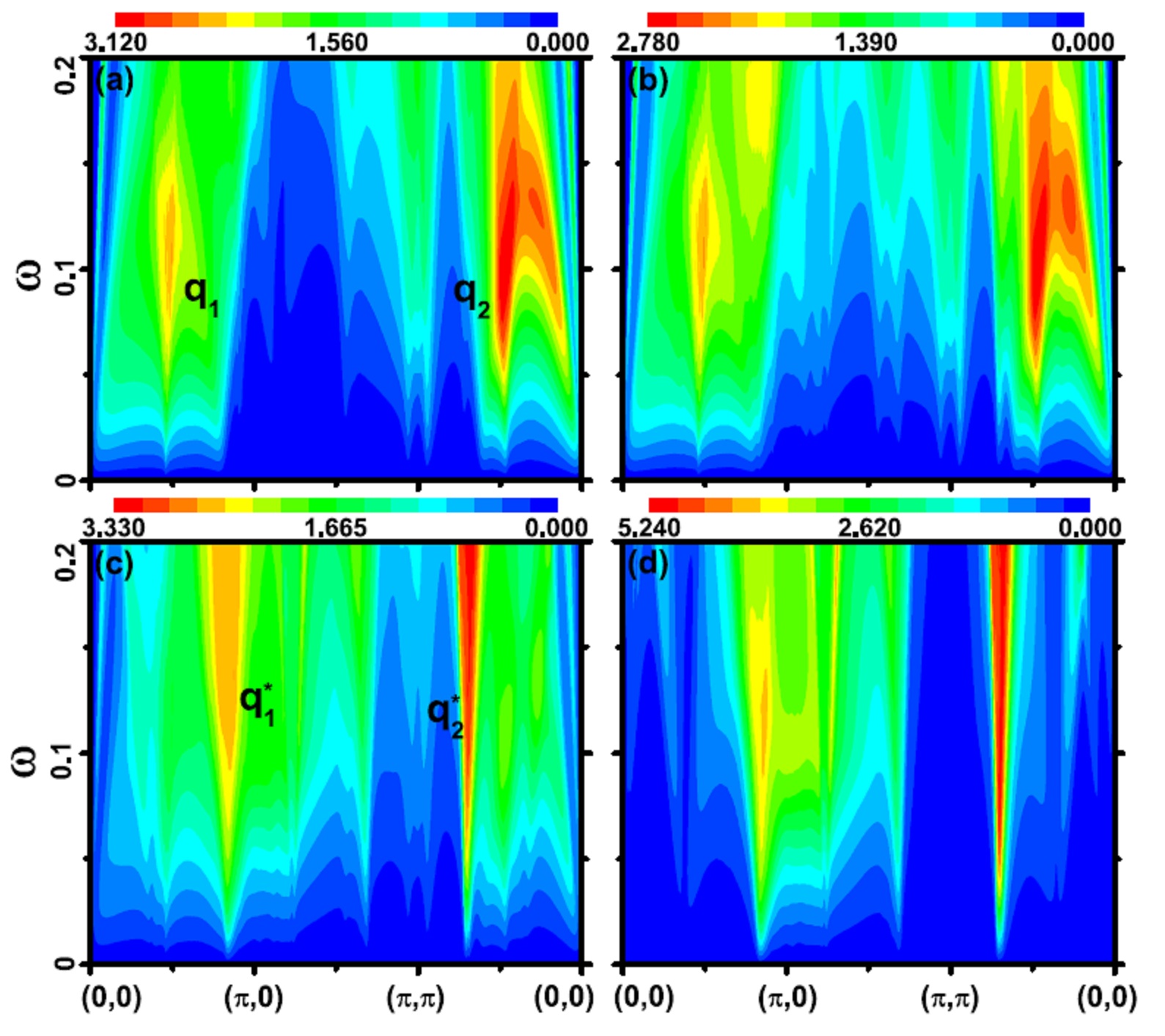}
	\caption{(Color online) The $q-\omega$ dependence of the MEs along the high symmetry direction. Panels (a), (b), (c) and (d) are for $q_{z}=0$, $\pi/4$, $\pi/2$ and $\pi$, respectively.}
	\label{MEdisp}
\end{figure}

To further reveal the $q_{z}$ evolution of the MEs. In Fig.~\ref{MEdisp}, we show their energy dependence along the high symmetry direction for various $q_{z}$. As depicted in panel (a) of Fig.~\ref{MEdisp}, for small $q_{z}$ near $0$, the MEs are dominated by the $q_1$ and $q_2$ modes which persist up to the high energy regime. With the increasing of $q_{z}$, the intensity of the $q_1$ and $q_2$ modes diminish gradually. Meanwhile, the $q_{1}^{*}$ and $q_{2}^{*}$ modes emerge with their intensity increasing steadily. They rule the behaviors of the MEs for large $q_{z}$. As shown in Fig.~\ref{MEomEVO} and \ref{MEdisp}, the MEs patterns of the nickelate La$_3$Ni$_2$O$_7$ change little with the increasing of energy. This is in sharp contrast with those of the copper-oxide~\cite{RPA2,RPA5} and iron-based superconductors~\cite{RPA13} where the structure transition of the MEs with energy usually happens. For cuprates in which the Van Hove singularity plays the crucial role in the behaviors of the MEs, the associated CECs move toward or away from the antinodal region with the variation of energy. Thus, a structure transition of the MEs with energy is usually expected~\cite{RPA2,RPA5}. While for iron-based superconductors, both the electron and the hole pockets are generally small in size. As a result, the shape of the CECs change significantly with energy, giving rise to the structure transition of the MEs~\cite{RPA13}. Here, we emphasize that the CECs of the energy bands of La$_3$Ni$_2$O$_7$ change little with energy, especially for those of the $\alpha$ and the $\beta$ pockets due to the fact that they are large in both the size and the Fermi velocity. As a result, the patterns of the MEs can persist up to the high energy regime. 

\subsection{Doping behaviors of the magnetic excitations around $n=3.0$}
\begin{figure}
	\centering\includegraphics[width=0.45\textwidth]{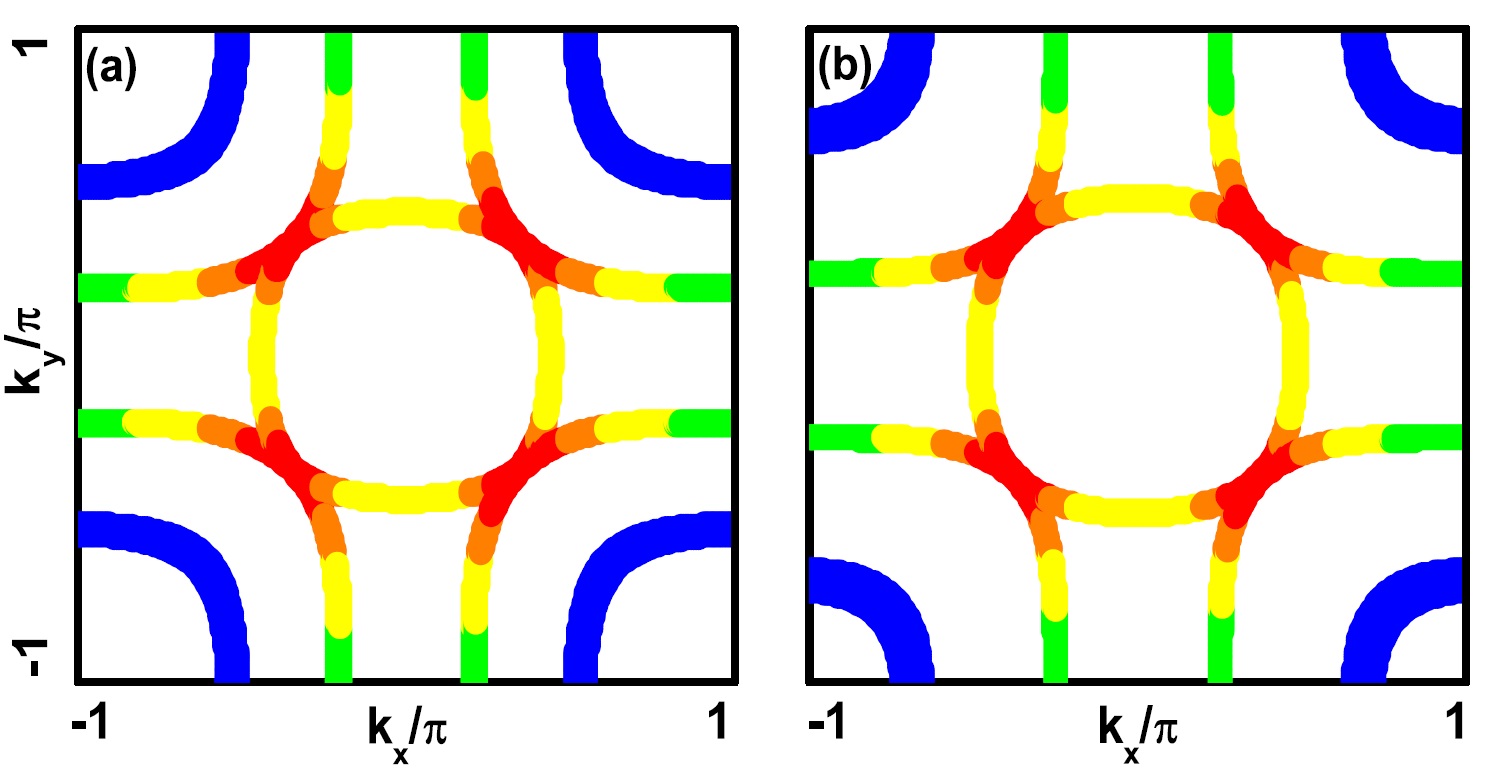}
	\caption{(Color online) The Fermi surfaces and their orbital compositions. Panels (a) and (b) are for $n=2.8$ and $3.2$, respectively.}
	\label{FSdoping}
\end{figure}

In the previous section, we discussed the normal state MEs with the filling number of $n=3.0$ which seems to be related to the moderately pressurized nickelate La$_3$Ni$_2$O$_7$ which exhibits a superconducting transition temperature of about $80$ K. It is interesting to know the doping evolution behaviors of the MEs because the topology of the Fermi surfaces may change prominently with carrier doping. In Fig.~\ref{FSdoping}, we plotted the typical Fermi surfaces and their orbital compositions for the hole-doped system of $n=2.8$ and the electron-doped one of $n=3.2$. Comparing to Fig.~\ref{U1.1OM0.02} (a), it can be seen that the $\alpha$ pocket shrinks with the decreasing of the filling number $n$. In contrast, the $\beta$ and $\gamma$ pockets expand steadily. Expect these, the topology of the Fermi surfaces keeps unchanged comparing to the case of $n=3.0$. 

\begin{figure}
	\centering\includegraphics[width=0.45\textwidth]{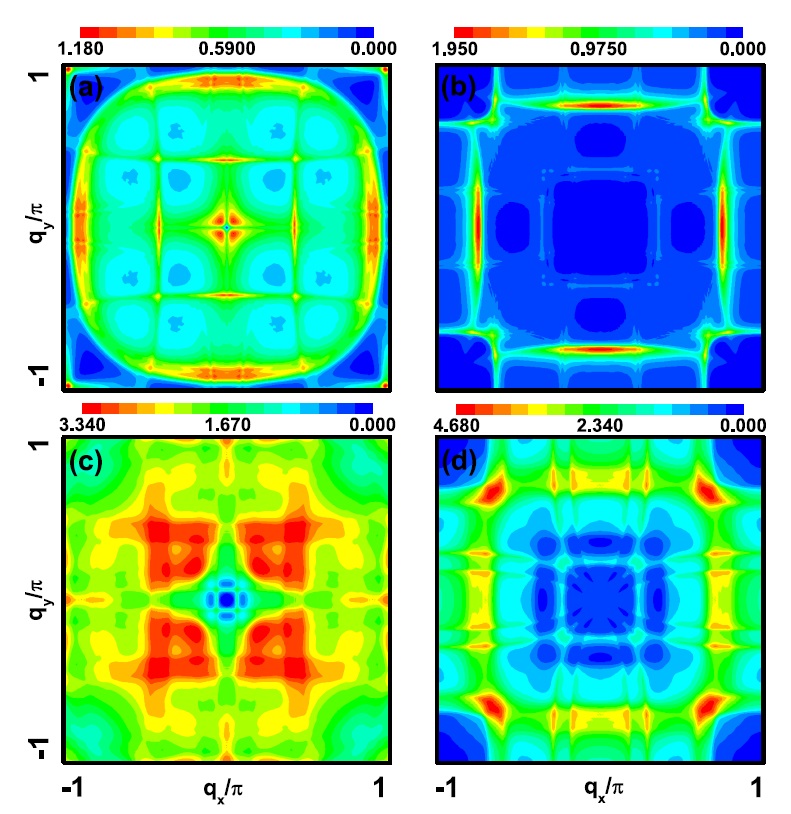}
	\caption{(Color online) The MEs for the hole-doped system with $n=2.8$. Panels (a) and (b) are for $\omega=0.02$ but with $q_{z}=0$ and $q_{z}=\pi$, respectively. Panels (c) and (d) are the same with (a) and (b) but for $\omega=0.14$.}
	\label{MEof2.8}
\end{figure}

In Fig.~\ref{MEof2.8}, we show the patterns of the MEs for the hole-doped system with $n=2.8$. As indicated by Fig.~\ref{MEof2.8} (a), the square-like MEs composed by the $q_1$ and $q_2$ modes and the FM correlations remain at a low energy of $\omega=0.02$ for $q_{z}=0$. In addition,  new MEs develop near $(0,\pm\pi)$ and $(\pm\pi,0)$ with sizable intensity. Actually, as shown in Fig.~\ref{U1.1OM0.02} (b), these MEs have already show up for $n=3.0$ where their intensity is weaker than that of the $q_1$ and $q_2$ modes. These MEs should be attributed to the particle-hole scatterings within the $\gamma$ pocket. It has been carefully checked that these modes move gradually toward $(0,\pm\pi)$ and $(\pm\pi,0)$ with the decreasing of the filling number. This conincides basically with the expansion of the $\gamma$ pocket with hole-doping. Together with the $q_{1}$ and $q_{2}$ modes, these MEs for small $q_{z}$ are all dominated by the intraband particle-hole excitations mainly from the $\gamma$ and the $\beta$ pockets, respectively. With the increasing of $q_{z}$, all these modes diminish gradually. Finally, the MEs are fully dominated by the $q_{1}^{*}$ and $q_{2}^{*}$ modes in the large $q_{z}$ regime, just like the $n=3.0$ case. It is interesting to notice that the $q_{1}^{*}$ modes move away from the $(0,\pm\pi)$ and $(\pm\pi,0)$ region. This is due to the fact that both the hole $\beta$ and $\gamma$ pockets expand in size with the decreasing of the filling number. Naturally, the length of $q_{1}^{*}$ decreases because it connects the $\beta$ pocket with the $\gamma$ one (see Fig.~\ref{U1.1OM0.02} (a)). This further confirms that the $q_{1}^{*}$ modes originate from the particle-hole scatterings between the $\beta$ and the $\gamma$ pockets. In contrast, the $q_{2}^{*}$ modes move rather slowly with hole-doping due to the much larger Fermi velocity of the $\alpha$ and $\beta$ bands comparing to that of the $\gamma$ one. The energy evolution behaviors of the MEs are quite similar to those presented in Fig.~\ref{MEomEVO} for $n=3.0$. With the increasing of energy, as shown in Fig.~\ref{MEof2.8} (c) for $\omega=0.14$, the typical MEs for small $q_{z}$ turn gradually to be dominated by the broad peaks around $q_{2}$. Furthermore, the $q_{z}$ evolutions of the MEs are also analogous to those of the $n=3.0$ case. Specifically, the $q_{2}^{*}$ and the splitted $q_{1}^{*}$ modes show up with prominent intensity for large $q_{z}$ around $\pi$. We have carefully checked that the above features of the MEs do not show qualitative changes up to the heavily hole-doped region around $n=2.6$. 

\begin{figure}
	\centering\includegraphics[width=0.45\textwidth]{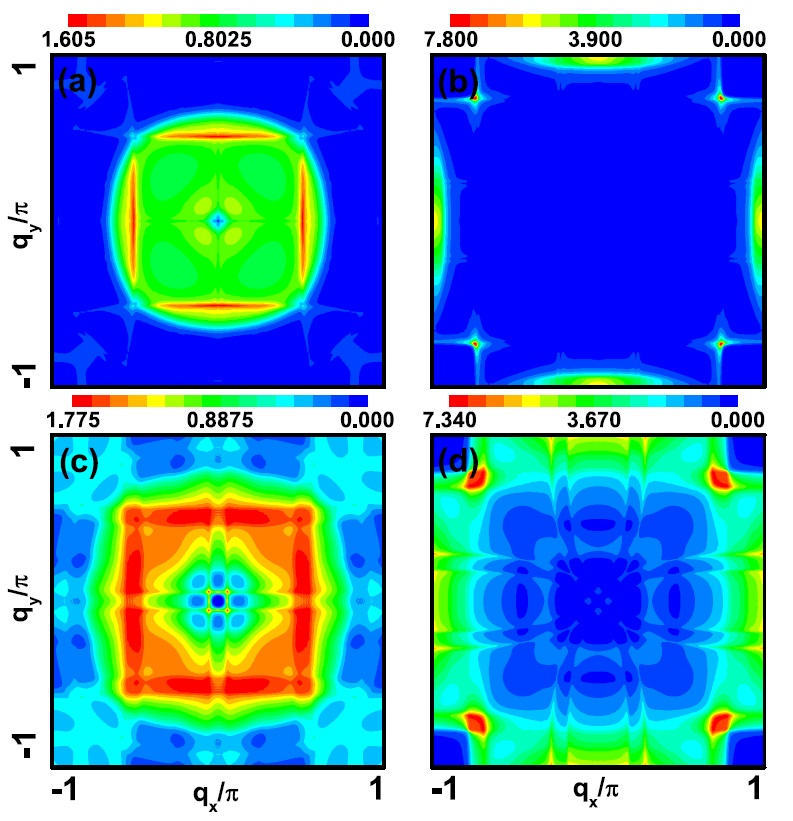}
	\caption{(Color online) The MEs for the electron-doped system with $n=3.2$. Panels (a) and (b) are for $\omega=0.02$ but with $q_{z}=0$ and $q_{z}=\pi$, respectively. Panels (c) and (d) are the same with (a) and (b) but for $\omega=0.14$.}
	\label{MEof3.2}
\end{figure}

For the electron-doped case, as shown in Fig.~\ref{MEof3.2}, the main features of the MEs remain for large $q_{z}$. They are also dominated by the $q_{1}^{*}$ and $q_{2}^{*}$ modes. Due to the contraction of the $\beta$ and the $\gamma$ pockets with electron-doping, the $q_{1}^{*}$ modes move toward the $(0,\pm\pi)$ and $(\pm\pi,0)$ region, contrary to that of the hole-doped case. And they reside just at $(0,\pm\pi)$ and $(\pm\pi,0)$ for $n=3.2$. Comparing to the hole-doped scenario, the low energy MEs near $(0,\pm\pi)$ and $(\pm\pi,0)$ disappear for small $q_{z}$, as shown in Fig.~\ref{MEof3.2} (a) for $q_{z}=0$. This is due to the fact that the hole $\gamma$ pocket shrinks and changes its shape from square to circle gradually with electron-doping. Thus, the original well-nesting property of the $\gamma$ pocket diminishes gradually, leading to the absence of the modes associated with the particle-hole excitations within the $\gamma$ pocket.

\subsection{Distinct MEs across the Lifshitz transition}

\begin{figure}
	\centering\includegraphics[width=0.45\textwidth]{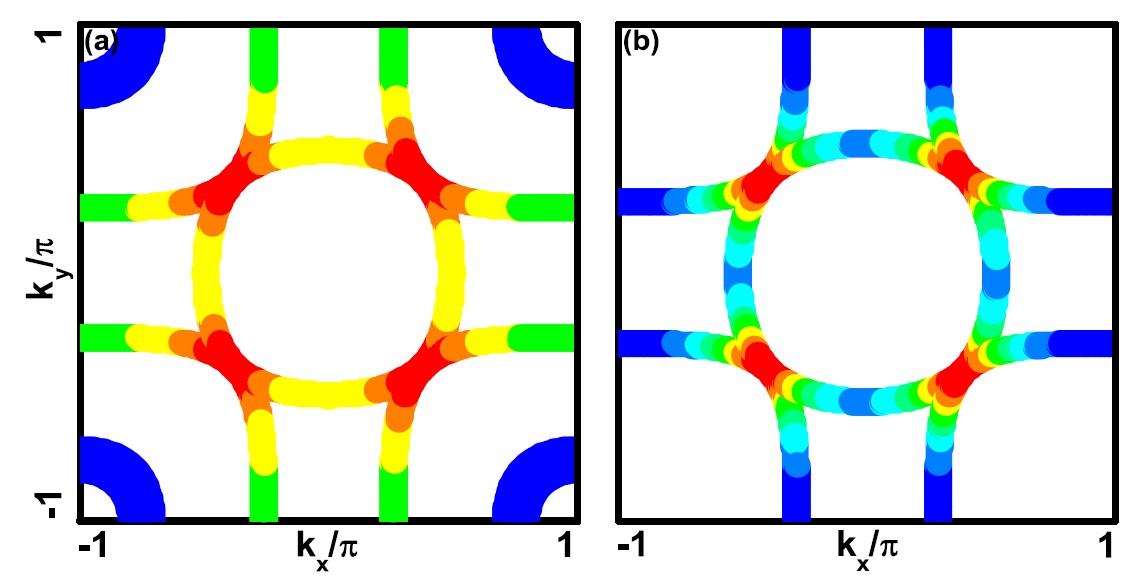}
	\caption{(Color online) The Fermi surfaces and their orbital compositions for $n=3.3$ (a) and $3.5$ (b), respectively.}
	\label{FSof3.3and3.5}
\end{figure}

With further increasing of electron-doping, the $\gamma$ pocket will turn gradually to be under the Fermi level, i.e., a Lifshitz transition will occur. This change of the Fermi surfaces topology is shown in Fig.~\ref{FSof3.3and3.5}. For the filling of $n=3.3$ above the Lifshitz transition, the hole $\gamma$ pocket coexists with the electron $\alpha$ and the hole $\beta$ ones. But when $n$ increases to be $3.5$, the $\gamma$ pocket disappears. Meanwhile, the orbital compositions of the $\alpha$ and the $\beta$ pockets change significantly around the Fermi level. In contrast to that above the Lifshitz transition where they are dominated by the $3d_{x^2-y^2}$ orbital. Below the transition, both pockets have prominent $3d_{z^{2}}$ orbital weight, especially around the $(\pm\pi,0)$ and $(0,\pm\pi)$ region. By studying the doping evolutions of the Fermi surfaces, it can be confirmed that the Lifshitz transition occurs around $n=3.4$. 

\begin{figure}
	\centering\includegraphics[width=0.45\textwidth]{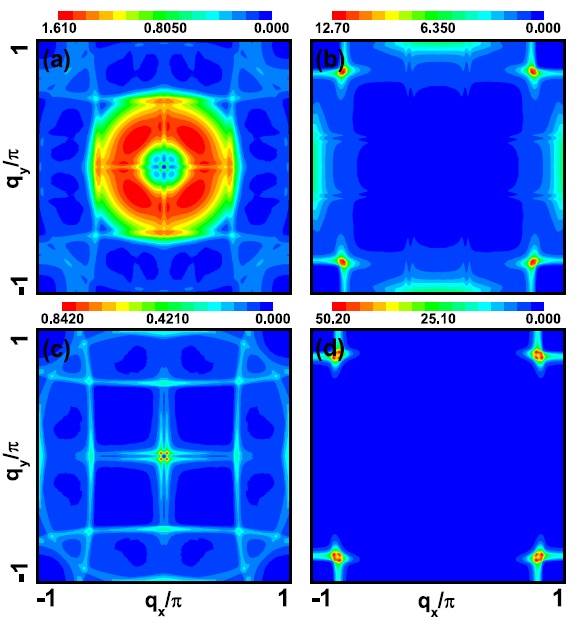}
	\caption{(Color online) The MEs across the Lifshitz transition around $n=3.4$. Panels (a) and (b) are the MEs for $n=3.3$, $\omega=0.06$ but with $q_{z}=0$ and $q_{z}=\pi$, respectively. Panels (c) and (d) are the same with (a) and (b) but for $n=3.5$.}
	\label{MEof3.3and3.5}
\end{figure}

Now we turn to the study of the MEs across the Lifshitz transition. For illustration, the typical MEs are shown in Fig.~\ref{MEof3.3and3.5} for $n=3.3$ and $3.5$ above and below the Lifshitz transition, respectively. For $n=3.3$, it can be seen from Fig.~\ref{MEof3.3and3.5} (a) that the $q_1$ and $q_2$ modes remain for $\omega=0.06$ and $q_{z}=0$, although the $q_2$ modes are rather weak in intensity. However, the most prominent feature of the MEs is a ring-like structure centered at $(0,0)$. This is fully attributed to the particle-hole scatterings within the small circular-shaped $\gamma$ pocket. Similar pattern of the MEs had been obtained for the monolayer cuprates~\cite{RPA14}. Thus, this kind of pattern can be seen as a hallmark of the MEs near the Lifshitz transition. With the increasing of $q_{z}$, the patterns of the MEs turn gradually to be dominated by the familiar $q_{1}^{*}$ and $q_{2}^{*}$ modes as usual. The typical MEs are shown in Fig.~\ref{MEof3.3and3.5} (b), they are nearly the same with those for $n=3.0$. Below the Lifshitz transition, the square-like MEs formed by the $q_{1}$ and $q_{2}$ modes remain for small $q_{z}$. This is due to the fact that these modes originate from the particle-hole scattterings between the flat segments of the $\beta$ pocket which changes little with electron-doping. In addition, as depicted in panel (c) of Fig.~\ref{MEof3.3and3.5}, the FM modes emerge near $(0,0)$. They should be attributed to the particle-hole scatterings between the $\alpha$ and the $\beta$ pockets around their touch points in the Brillouin zone. We have checked that these FM modes move away from $(0,0)$ gradually with the increasing of energy due to the splitting between the CECs of the $\alpha$ and the $\beta$ pockets. With the increasing of $q_{z}$, the patterns of the MEs change dramatically. For large $q_{z}$ around $\pi$, the $q_{1}^{*}$ modes disappear totally due to the absence of the $\gamma$ pocket, and the resultant MEs are fully dominated by the $q_{2}^{*}$ modes with rather high intensity which indicates a strong tendency toward the SDW order. It should be noticed that, for this SDW order, the in-plane spins tend to form an incommensurate order with momentum $q_{2}^{*}$ near $(\pi,\pi)$, while they prefer to align antiferromagnetically between the upper and the lower nickel-oxygen layers.

Actually, we have performed the detailed numerical calculations of the MEs for the wide doping region from $n=3.4$ to $n=4.0$. We find that the above mentioned features of the MEs are very robust below the Lifshitz transition. Both the square-like pattern of the MEs for small $q_{z}$ and the $q_{2}^{*}$ modes for large $q_{z}$ persist up to the high energy regime. Furthermore, all these behaviors of the MEs remain in the wide doping region from $n=3.5$ to $n=4.0$. Moreover, the intensity of $q_{2}^{*}$ modes overwhelms that with small $q_{z}$ in this doping region. And it grows rapidly with electron-doping which signals the strong tendency toward the magnetic order. This is in sharp contrast with the MEs above the Lifshitz transition where intensities of the small $q_{z}$ modes are usually comparable to those with large $q_{z}$. The strong tendency toward the SDW order can be understood in the following way. Below the Lifshitz transition, the $\gamma$ band is fully occupied with the electron number $n_{\gamma}=2$. For a filling of $n=4$, the size of the electron $\alpha$ and the hole $\beta$ pockets should be equal to fulfill $n_{\alpha}+n_{\beta}=2$. Considering the square-like shape of the Fermi surfaces, there should exist perfect nesting between the electron $\alpha$ and the hole $\beta$ pockets, just like what happened in the iron-based superconductors~\cite{Mazin}. Thus, one can expect that the pressurized La$_3$Ni$_2$O$_7$ shall be in the SDW phase which ordered at $(\pi,\pi,\pi)$ around the filing of $n=4$. In this way, it seems that the superconducting phase is in close proximity to the SDW state and it can emerge through hole-doping to the "parental compound" of the pressurized nickelates with $n=4$. From this point of view, the pressurized nickel-based superconductors seem to share some common features with their iron-based counterparts.

\section{Summary and conclusions}
In summary, we have studied the normal state behaviors of the MEs in the double-layer pressurized nickelate La$_3$Ni$_2$O$_7$ based on a realistic theoretical model~\cite{Yaotbmodel}. For $n=3.0$ which seems to be most relevant to the material which realizes the high temperature superconductivity of about $80$ K under the moderate pressure above $14$ GPa. It was found that the MEs have very strong interlayer modulations by $q_{z}$ due to the displacement between the upper and the lower nickel-oxygen layers. For small $q_{z}$, the low energy MEs are dominated by a square-like structure composed of the $q_1$ and $q_2$ modes which both originate from the intraband particle-hole scatterings of the $\beta$ pocket. The square-shaped patterns of the MEs persist with the increasing of energy, but they turn to be four broad peaks around $(\pm\pi/2, \pm\pi/2)$ in the high energy regime. The pattern of the MEs changes dramatically when $q_{z}$ becomes large. Especially, the square pattern of the MEs for small $q_{z}$ disappears gradually with the increasing of  $q_{z}$. Finally, for large $q_{z}$ around $\pi$, the MEs are dominated by the $q_{1}^{*}$ modes which locate around $(\pm\pi, 0)$ and $(0,\pm\pi)$ and the $q_{2}^{*}$ modes around $(\pi,\pi)$. To our knowledge, these hidden $q_{z}$ dependent structures of the MEs had not been established by previous studies~\cite{Yaotbmodel,EreminRPAandpairing}. It was carefully checked that the $q_{1}^{*}$ modes originate from the particle-hole scatterings between the $\beta$ and the $\gamma$ pockets. While the $q_{2}^{*}$ modes mainly arise from the particle-hole excitations between the $\beta$ and the $\alpha$ pockets, and they split gradually into parts in the high energy regime. Generally, the MEs show significantly larger intensity for the modes with $q_{z}$ near $\pi$. Thus, these interlayer antiferromagnetic modes are expected to play a prominent role in the formation of cooper pairs. Actually, the patterns of the MEs found here for large $q_{z}$ are very similar to the momentum structure of the SDW interaction established before which was proposed to be the pairing glue~\cite{WangqianghuaPrb}.  

With hole-doping, both the square pattern of the MEs for small $q_{z}$ and the $q_{1}^{*}$ and $q_{2}^{*}$ modes for large $q_{z}$ persist up to the heavily hole-doping regime around $n=2.6$, although new MEs will emerge near the boundary of the Brillouin zone attributing to the particle-hole scatterings within the $\gamma$ pocket which turns to be the square-like shape with the gradually increasing of hole-doping. On the other hand, for eletron-doping up to $n=3.3$, the patterns of the MEs and their energy and $q_{z}$ dependence are qualitatively the same with those for $n=3.0$. Thus, the MEs exhibit very robust features in both the hole- and electron-doping region around $n=3.0$. 

With further increasing of electron-doping, the $\gamma$ pocket turns gradually to be under the Fermi level, i.e., a Lifshitz transition will occur around $n=3.4$. Consequently, the behaviors of the MEs change dramatically. Just above the transition, the MEs for small $q_{z}$ turn to be dominated by a ring-like structure centered at $(0,0)$ instead of the square-like pattern far above the transition point. These ring-like MEs should be attributed to the particle-hole scatterings within the $\gamma$ pocket. Such behavior of the MEs had been established for the monolayer cuprates before~\cite{RPA14}. Thus, it seems to be a hallmark of the low energy MEs near the Lifshitz transition. In contrast, the MEs for large $q_{z}$ are still dominated by the $q_{1}^{*}$ and $q_{2}^{*}$ modes, although the intensity of the $q_{1}^{*}$ modes weakens significantly. Below the transition, the behaviors of the MEs change dramatically. The most prominent features of the MEs for large $q_{z}$ are the absence of the $q_{1}^{*}$ modes and the strongly enhanced intensity of the $q_{2}^{*}$ modes. In contrast, the MEs with small $q_{z}$ keep to be the square-like pattern but with much weakened intensity. As a result, the MEs are fully ruled by the $q_{2}^{*}$ modes which indicate the strong tendency toward the SDW order. 

It is interesting to notice that the $\mu$SR~\cite{uSR,uSRsplitDW}, NMR~\cite{ChenxianhuiNMR} and RIXS~\cite{FengRIXS} measurements on the double-layer nickelates had confirmed the development of the SDW order at ambient pressure. Furthermore, it was found that the SDW ordered at $(\pi/2,\pi/2,\pi)$. Thus, the superconductivity is in close proximity with the SDW state in the pressure-temperature phase diagram. However, here we investigate the doping evolution behaviors of the MEs of the double-layer nickelate La$_3$Ni$_2$O$_7$ with fixed moderate pressure at which superconductivity emerges. We predict the existence of a SDW phase which orders around $(\pi,\pi,\pi)$ for the heavily electron-doped material under moderate pressures. This SDW instability is governed by the perfect nesting between the $\alpha$ and the $\beta$ pockets at $n=4.0$. Due to the fact that the spin susceptibility will be further enhanced by interaction, the SDW state may develop around $n=3.5$, i.e., the doping level of about $0.25$ electron per Ni-site, according to our numerical results. If this is true, it seems that superconductivity emerges through hole-doping to the "parental compound" of the pressurized nickelates with $n=4.0$. In this way, the nickel-based superconductors share very similar features with their iron-based counterparts. We expect the doping-temperature phase diagram of the pressurized La$_3$Ni$_2$O$_7$ will be explored experimentally and theoretically in the near future.

\section{Acknowledgments}
This work was supported by the National Natural Science Foundation of China (Grants Nos.~11804290, 11647072).

\end{document}